# Why the dark matter of galaxies is clumps of micro-brown-dwarfs and not Cold Dark Matter

Carl H. Gibson[1,2]

[1]University of California San Diego, La Jolla, CA 92093-0411, USA

[2]cgibson@ucsd.edu, http://sdcc3.ucsd.edu/~ir118

**ABSTRACT**

Observations of quasar microlensing by Schild 1996 show the baryonic dark matter BDM of galaxies is micro-brown-dwarfs, primordial hydrogen-helium planets formed at the plasma to gas transition $10^{13}$ seconds, in trillion-planet clumps termed proto-globular-star-clusters PGCs.  Large photon-viscosity ν of the plasma permits supercluster-mass gravitational fragmentation at $10^{12}$ seconds when the horizon scale $L_H = ct$ is matched by the Schwarz viscous scale $L_{SV}$ of Gibson 1996.  Voids begin expansion at sonic speeds $c/3^{1/2}$, where $c$ is light speed and $t$ is time, explaining $10^{25}$ meter size regions observed to be devoid of all matter, either BDM or non-baryonic NBDM.  Most of the NBDM is weakly-collisional, strongly-diffusive, neutrino-like particles.  If cold NBDM (CDM) is assumed, it must soon become warm and diffuse because it is weakly-collisional.  It cannot clump and its clumps cannot clump.  CDM is ruled out with 99% confidence by local–group satellite observations of Kroupa et al. 2010.  The satellites are clusters of PGCs.  PGCs are recaptured by the Galaxy on an accretion disk as they freeze and diffuse from its core to form its BDM halo.  Stars form by viscous mergers of primordial gas planets within PGCs.  Stars die by overeating mBDs, making the first chemicals, oceans and life at 2-8 Myr.

## 1. Introduction

Dark matter is missing mass needed to explain why rapidly rotating objects like galaxies do not fly apart due to centrifugal forces.  The candidates for the dark matter are ordinary baryonic matter (protons etc.), non-baryonic matter (like neutrinos) and dark energy (a permanent anti-gravity material).  According to the standard concordance model of cosmology ΛCDMHC, 70% of the missing mass is dark energy and about 4% is baryonic. The rest is a mix of non-baryonic dark matter NBDM neutrinos (Nieuwenhuizen 2009, 2011) and a mysterious material called "cold dark matter" CDM. The collisional properties of the baryonic dark matter are crucial to and determine the gravitational structure formation of both the plasma and gas epochs (Nieuwenhuizen et al. 2009, 2010ab, 2011, Gibson & Schild 2010, Schild & Gibson 2010). CDM is a myth that is easily dispelled by hydrogravitational dynamics HGD theory; that is, proper inclusion of basic ideas of fluid mechanics.

HGD modification of the standard model to include fluid mechanical effects of viscosity, diffusivity, fossil turbulence and turbulence (Gibson 1996, 2000, 2004, 2005) greatly simplifies the theory of gravitational structure formation and explains a large variety of observation mysteries, including the big bang itself (Gibson 2010, Gibson, Schild & Wickramasinghe 2011). Dark energy and cold dark matter are not required to explain cosmology, and are in serious conflict with a wide variety of observations. Consider the increase in the "missing mass" to luminosity ratio M/L as a function of averaging scale shown in Figure 1 (Silk 1994 p124). The missing mass M supplies gravitational attraction to prevent spinning luminous objects like galaxies and galaxy clusters from flying apart because the luminous stars have insufficient mass to overcome estimates of centrifugal forces.

Dynamical mass increases relative to luminous mass as observations extend to larger and larger scales using a variety of dynamical mass measures (Zwicky 1937). HGD analysis explains star formation, planet formation and how stars die by excessive accretion of planets till the stars exceed critical mass values and explode as supernovae. Chemicals of supernovae produce the first water oceans and a biological big bang at only 2-8 Myr when cometary panspermia of Hoyle-Wickramasinghe is merged with HGD cosmology (Gibson, Wickramasinghe and Schild 2010).

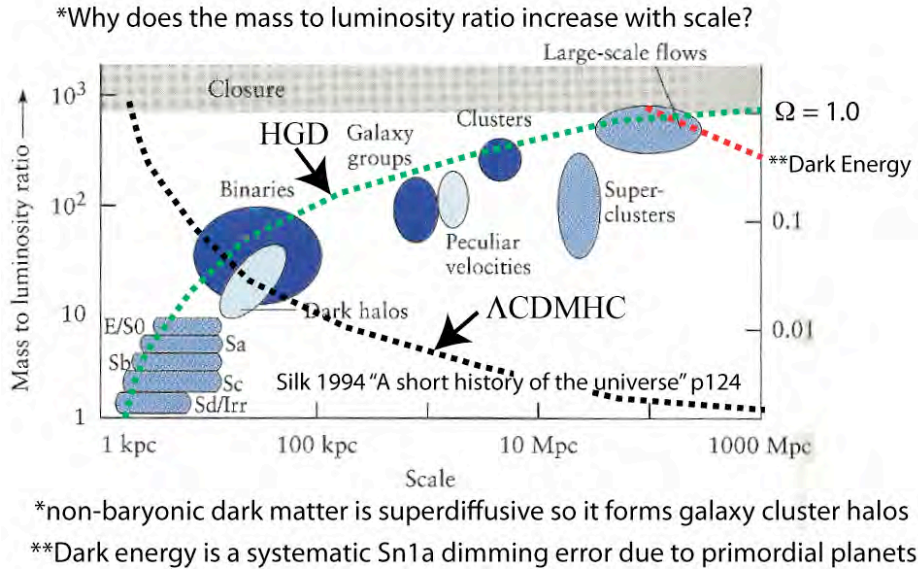

Fig. 1.  Missing mass to luminosity M/L increases with scale due to the diffusivity of the dark matter, from a factor of about 30 for galaxies, to ~ $10^3$ for superclusters of galaxies (Silk 1994). According to hydrogravitational dynamics theory HGD (Gibson 1996) the weakly collisional non-baryonic dark matter NBDM is superdiffusive.  It forms halos for superclusters.  Antigravitational forces of turbulence produce the big bang but they dissipate and are not permanent like $\Lambda$.

Only about a factor M/L of thirty of the mass is missing for galaxies, from their flat rotation curves that extend to about 30 kpc ($10^{21}$ meters).  However, considerably more M/L ~ $10^3$ is needed to stabilize galaxy clusters and superclusters.  If CDM clumps were really stable (they are not) and the clumps of CDM could really cluster (they cannot) then the density of the missing mass to luminosity ratio M/L should decrease with averaging length scale (not increase as usually assumed).  This is shown in Fig. 1 (black dotted line), starting from infinity before the appearance of the first star where L = 0 and M is whatever quantity of CDM clumps are needed to gravitationally collect enough baryons to produce the first star, and approaching zero at large averaging scales where a few massive CDM halos become very bright as they fill with turbulent superstars and supernovae that reionize the universe.

Reionization of the universe is unnecessary according to HGD theory because the missing hydrogen of the Lyman-$\alpha$ forest is sequestered by primordial planets. Another unlikely standard assumption is that most of the baryons are in the intergalactic medium (Madau 2005).  This follows if galaxies are formed in CDM halos but does not if galaxies form by fragmentation of the plasma. From HGD, the mechanisms for gas to enter the sparse intergalactic medium are by AGN jets during the brief period of central black hole formation, and by supernovae, but careful observations show at least 90% of the baryons are missing from both galaxies and the intergalactic medium at small redshifts (Bregman 2007). X-ray gas at large z suggests a massive Warm-Hot Intergalactic Medium WHIM for $z \sim 0$ to account for the missing baryons, but fails as expected from HGD.

Predictions of HGD cosmology (green dotted) and Dark Energy (red dotted) curves are compared to the observations in Fig. 1. According to HGD, the non-baryonic dark matter cannot condense because it is weakly collisional, however cold it is assumed to be. The diffusivity of a material is proportional to the mean free path for particle collisions times the speed of the material particles, and this increases as the particle collision cross-section decreases. Therefore the reason the observed M/L ratio increases with scale in Fig. 1 is that the non-baryonic dark matter NBDM is superdiffusive because its particles are weakly collisional. NBDM diffuses to $\sim 10^3$ larger scales than the 30 kpc observed for galaxies ($10^{22}$ m), to the 300 Mpc size of supercluster halos ($10^{25}$ m).

Kroupa et al. (2010) test predictions of ΛCDM models against observations of the local group of Milky Way satellites and conclude the "concordance" ΛCDMHC model is unacceptable, as shown in Figure 2. From model clustering properties assigned to CDM clumps, there should be large numbers of such clumps in the Milky Way, but these are not observed, as shown in Fig. 2A. New local group satellites termed dwarf spheroidals are detected with large M/L values $\sim 10^3$, Fig. 2C. Instead of being randomly scattered in a dark matter halo as expected from ΛCDMHC, the new faint satellites (green circles) and previously known satellites (yellow circles) are assembled in a "disk of satellites" DoS structure (predicted by HGD due to the sticky nature of PGC dark matter diffusing from the central protogalaxy core on freezing).

HGD cosmology predicts a PGC accretion disk, or "disk of structures" DoS structure, where dense plasma protogalaxies PGs are the least massive structures to form during the plasma epoch. The most massive were superclusters starting at $10^{12}$ seconds when plasma viscous forces first permitted fragmentation ($L_{SV} \sim L_H$, Table 1). PGs fragment into protoglobularstarcluster PGC clumps of primordial planets PFPs at the time of plasma to gas transition $10^{13}$ seconds, and diffuse to form the baryonic dark matter halo and the DoS accretion disk as the planets gradually freeze and the diffusivity of the PGC clumps of planets increases so the dark matter halo can form by diffusion of the baryonic dark matter BDM.

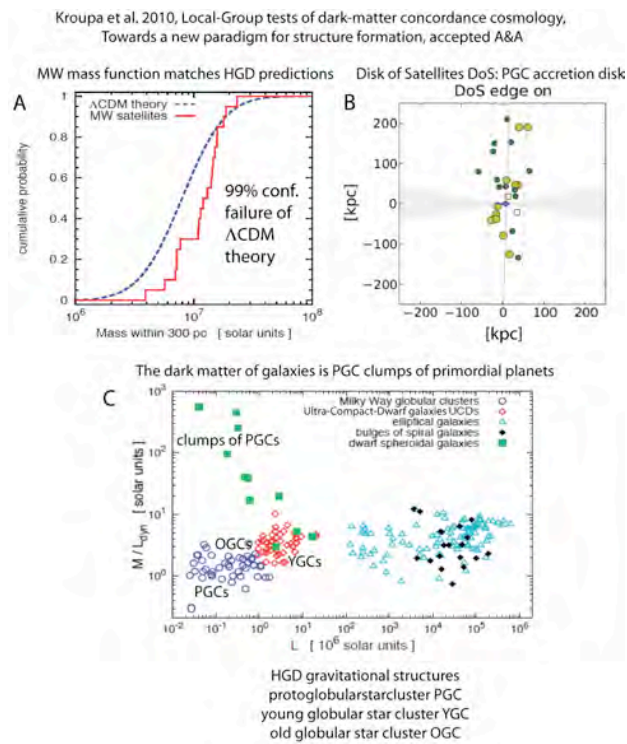

Fig. 2. Kroupa et al. (2010) evidence of ΛCDM failures to describe the local group of Milky Way satellites. A. The mass function of ΛCDM fails badly. B. The new satellites (green) are extremely dim and clustered in the known (yellow) disk of satellites (DoS). C. The M/L ratio for gravitational structures supports HGD predictions (bottom).

Cold dark matter was invented to accommodate the Jeans acoustic criterion for gravitational structure formation, where the Jeans length scale $L_J = (V_S/\rho G)^{1/2}$. Because the sound speed of the plasma is so large, its Jeans scale exceeds the scale of causal connection $ct$ prior to decoupling. In the standard cosmological model, in order to produce structure, it was assumed that the non-baryonic dark matter was cold to reduce its sound speed and permit condensations. We suggest this idea is physically impossible. A clump of CDM will contract by gravitational forces until its density permits collisions. It will then proceed to diffuse away as the gravitational potential energy of the CDM clump distributes itself. Clumps of CDM cannot merge to larger and larger clumps by hierarchical clustering HC. CDM clumps of all sizes are diffusively unstable.

A permanent dark energy $\Lambda$ component in the universe is also unlikely and unnecessary. Such a material would cause an accelerated expansion of the universe and a decrease of density below that necessary for the universe to be closed (Fig. 1 red dotted line). According to HGD the initial big bang event is due to a turbulent instability (Gibson 2004, 2005). Because turbulence is a dissipative process, the universe must be closed (Fig. 1 green dotted line).

## 2. Fluid mechanics of the early universe

Figure 3 summarizes the evolution of gravitational structures in the universe according to hydrogravitational dynamics HGD theory. Table 1 summarizes the various critical length scales that enter the analysis, Gibson (1996).

Table 1. Length scales of gravitational instability

| Length Scale Name | Definition | Physical Significance |
|---|---|---|
| Jeans Acoustic | $L_J = V_S/(\rho G)^{1/2}$ | Acoustic time matches free fall time |
| Schwarz Viscous | $L_{SV} = (\gamma \nu/\rho G)^{1/2}$ | Viscous forces match gravitational forces |
| Schwarz Turbulent | $L_{ST} = (\varepsilon/[\rho G]^{3/2})^{1/2}$ | Turbulent forces match gravitational forces |
| Schwarz Diffusive | $L_{SD} = (D^2/\rho G)^{1/4}$ | Diffusive speed matches free fall speed |
| Horizon, causal connection | $L_H = ct$ | Range of possible gravitational interaction |
| Plummer force scale | $L_{CDM}$ | Artificial numerical CDM halo sticking length |

$V_S$ is sound speed, $\rho$ is density, G is Newton's constant, $\gamma$ is the rate of strain, $\nu$ is the kinematic viscosity, $\varepsilon$ is viscous dissipation rate, D is the diffusivity, $c$ is light speed, $t$ is time.

On the right of Fig. 3 is the turbulent big bang (Gibson 2004, 2005). When the local temperature exceeds the Planck temperature of $10^{32}$ K, Planck mass particles and anti-particles with mass $10^{-8}$ kg can spontaneously appear from the vacuum and return to the vacuum without effect within a Planck time $10^{-43}$ seconds. However, rather than annihilation, the pair can merge to form the Planck equivalent of positronium, formed by electrons and positrons in supernovae. A prograde merger of an anti-Planck particle with such a Planck particle pair results in the HGD turbulent big bang universe. Further details and supporting evidence are provided by Gibson (2010) and Gibson and Schild (2010).

Strong antigravitational forces are required to power the big bang. At first these are supplied by inertial vortex forces of the big bang turbulence fireball. When the expanding, cooling fireball reaches $10^{28}$ K a phase change occurs. Quarks and gluons become possible. Negative pressures from gluon viscosity work against the expansion with power $10^{145}$ watts, producing the $\sim 10^{90}$ kg of the present universe.

The magnitude of big bang pressures is the Fortov-Planck pressure $L_{FP} = c^7 h^{-1} G^{-2}$ of 4.6 $10^{113}$ Pa (Keeler and Gibson 2009), where c is light speed, h is Planck's constant and G is Newton's constant (Fortov 2009). Inertial vortex forces of big bang turbulence produce negative pressures larger than $L_{FP}$ during the initial stages of the spinning, turbulent, big bang fireball in order to extract mass-energy from the vacuum following Einstein's equations (see Peacock 2000, p. 19, section 1.5). Much larger negative stresses and mass-energy production occur during inflation once the quark-gluon plasma appears producing gluon viscosity.

The Schwarz diffusive scale $L_{SD}$ of the non-baryonic dark matter is larger than the horizon scale $L_H$ throughout the plasma epoch, so NBDM is taken to be a passive secondary factor in the gravitational structure formation of clusters and galaxies by the plasma according to HGD theory. Supercluster halos form at $L_{SD}$ scales $\sim 3 \times 10^{22}$ meters at time $\sim 10^{14}$ s soon after decoupling by fragmentation at the scale of causal connection, matching observations. The indicated diffusivity of the NBDM is $\sim 10^{30}$ m$^2$ s$^{-1}$ (Gibson 2000).

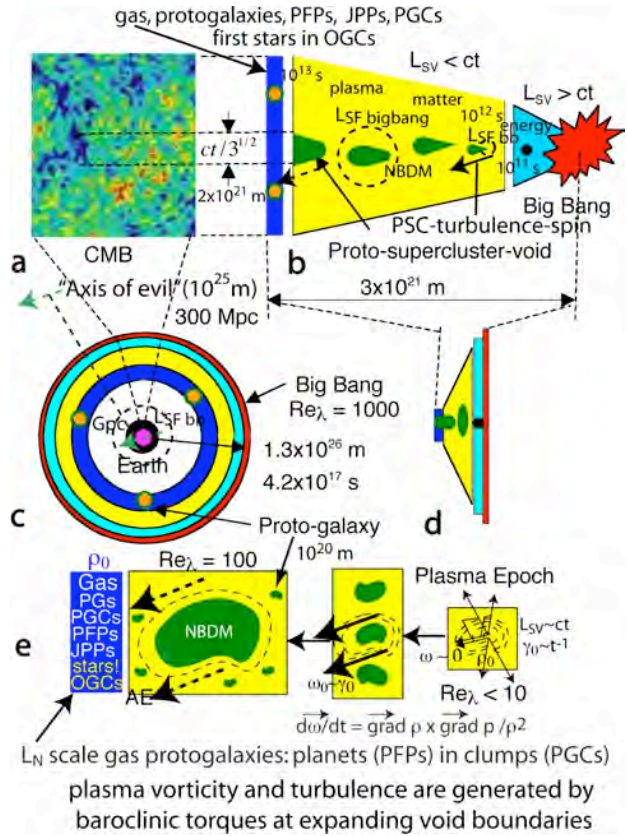

Fig. 3. Hydrogravitational dynamics HGD model for the development of structure in the universe.

Viscous forces control the first formation of gravitational structures during the plasma epoch beginning at time $10^{12}$ seconds (30,000 years) after the big bang. It is physically impossible for the non-baryonic dark matter to condense to form permanent clumps, as assumed by the ΛCDMHC standard model of cosmology. This is illustrated in Figure 4. In Fig. 3b we see the first structure is due to viscous-gravitational instability of the baryonic plasma at density minima, leading to fragmentation of proto-supercluster-voids that expand to scales detected in CMB temperature anomalies of Fig. 3a.

Turbulence forces and turbulence morphology develop during the plasma epoch, as shown in Fig. 3e. Our local fossil turbulence vortex line from the big bang is manifested by the "Axis of Evil", Schild and Gibson (2008). Vorticity is generated by baroclinic torques at expanding PSC-void boundaries to produce weak turbulence, whose fossil turbulence vortex lines trigger the formation of chains of proto-galaxies at the end of the plasma epoch.

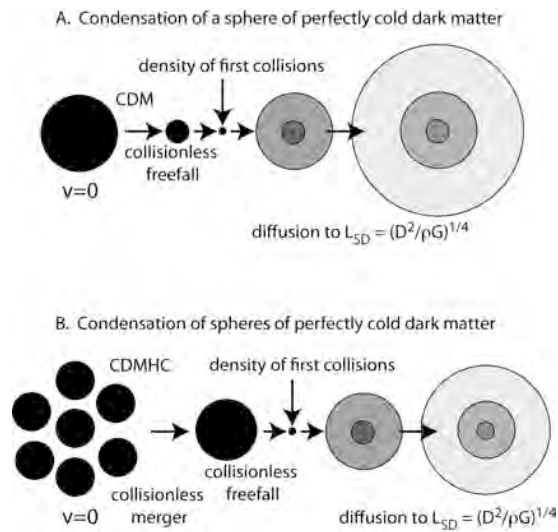

Fig. 4. Problems with the standard cosmological model of structure formation begin with the assumption that cold dark matter can remain in stable clumps.  It is physically impossible for this to happen.  Starting from v=0 the CDM clump in A. will experience a collisionless freefall to a smaller size and larger density where collisions will begin.  The smaller the collision cross section the smaller the size and the larger the density.  The material will then diffuse to scale $L_{SD}$ (Gibson 2000).  Multiple clumps in B. will evolve in a similar fashion.

As shown in Fig. 4 A, cold dark matter CDM clumps are unstable to gravitational collapse and diffusion independent of the collision cross section assumed for the non-baryonic-dark-matter NBDM particles, whatever they are.  As shown in Fig. 4 B, hierarchically clustering HC clumps of CDM clumps are equally unstable to clumping and clustering and cannot possibly form the necessary "CDM halos" of concordance cosmology to collect baryons by forming deep gravitational potential wells.

The CDMHC mechanism is physically unrealistic, and should be abandoned.  Numerous numerical simulations have been published that claim support for the CDMHC mechanism.  They are not to be trusted because they ignore problems of diffusivity of NBDM illustrated in Fig. 4AB and assume an N body problem of gravitationally interacting objects more appropriate to proto-globular-star-cluster PGC clumps of primordial planets in hydrogravitational dynamics HGD theory than CDM halos (green dotted curve of Fig. 1). Merging of CDM clumps to form larger stable clumps is not represented by N body gravitational simulations as assumed. A proper, logically consistent, numerical simulation of CDMHC behavior in Fig. 1 would show behavior quite orthogonal (black dotted curve of Fig. 1) to that expected for the universe where structure formation is guided by HGD theory (green dotted curve of Fig. 1).

## 3. Evidence that the dark matter of galaxies is primordial planets in clumps

The best evidence that the mass of galaxies is planet mass objects is the sequence of quasar microlensing studies discussed by Schild (1996), Colley and Schild (2003) and Colley et al. (2003). A foreground galaxy collects light from a background quasar on its line of sight. Light curves from mirage images are subtracted after shifting by the difference in their time delays to determine twinkling frequencies of the approximately point mass objects doing the microlensing. With the extremely precise time delay between the A and B images of 417.1 days, a small planetary quasar-crossing event time of only 12 hours was detected (Colley and Schild 2003).

For the Schild-Quasar, carefully studied for 15 years, and then followed up by the world-wide, round-the-clock, collaboration of Colley et al. (2003) giving precise $10^{-9}$ arc-second resolution of the background quasar, it becomes abundantly clear that the dominant galaxy-mass objects are not stars but planets. The twinkling times were not years but days and even hours. Because the mirage images differed slightly in brightness they were not uniformly distributed, but clumped (Gibson & Schild 2010, this volume). Collaborations (MACHO, EROS, etc.) that failed to detect planetary mass objects as the dark matter of galaxies have all made the fatally flawed assumption that the objects are not clumped but uniformly distributed.

Further evidence that most of the mass of galaxies is frozen hydrogen planets is given by temperatures detected by infrared and microwave space telescopes, as shown in Figure 5. Fig. 5 (top) shows the Herschel space telescope and other "dust" temperature estimates as a function of redshift back to z > 1.3. Remarkably, the temperatures are close to what one would expect if the "dust" is not the talcum powder ceramic variety detected in comets but more likely frozen hydrogen planets in the process of star formation as predicted by HGD cosmology. The triple point of hydrogen is 13.8 K, and this is the observed lower bound. The critical point of hydrogen is 32 K, close to the upper bound of the observed temperatures. These are the thermostated temperatures one would expect for baryonic dark matter planets in clumps where the planets are merging to form stars and the stars are heating the planets to form large atmospheres of evaporated hydrogen and helium exposed to and in radiative equilibrium with the lower temperatures of outer space (green dotted line).

The ultra-luminous-infrared-galaxies (red boxes) are at temperatures above the 13.8-32 K hydrogen thermostat band. The implication is that such galaxies are particularly well supplied with planets rich in polycyclic-aromatic-hydrocarbon PAH dust. Evaporated planets and their PAH rich atmospheres are re-radiating infrared energy absorbed by such planets surrounding an active galactic nucleus that would otherwise appear in the optical frequency band (Gibson 2010, Gibson, Schild & Wickramasinghe 2010). This process will be discussed further in future work.

Dunne et al. (2011) show a rapid evolution of the "dust" of Fig. 5 over the last 5 billion years, also from the Herschel-ATLAS data. The temperature of the "dust" from nearly 2000 galaxies is binned according to temperature in figure 7 (Dunne et al. 2011 fig 7) and shows two remarkable spikes centered on 20 K and 45 K. This observation is easily explained from HGD cosmology as a reflection of the freezing-evaporation equilibrium of hydrogen dark matter planets in PGC clumps at 20K, and the expected hydrogen dark matter planet critical-boiling point temperature at 45 K near hot and exploding stars.

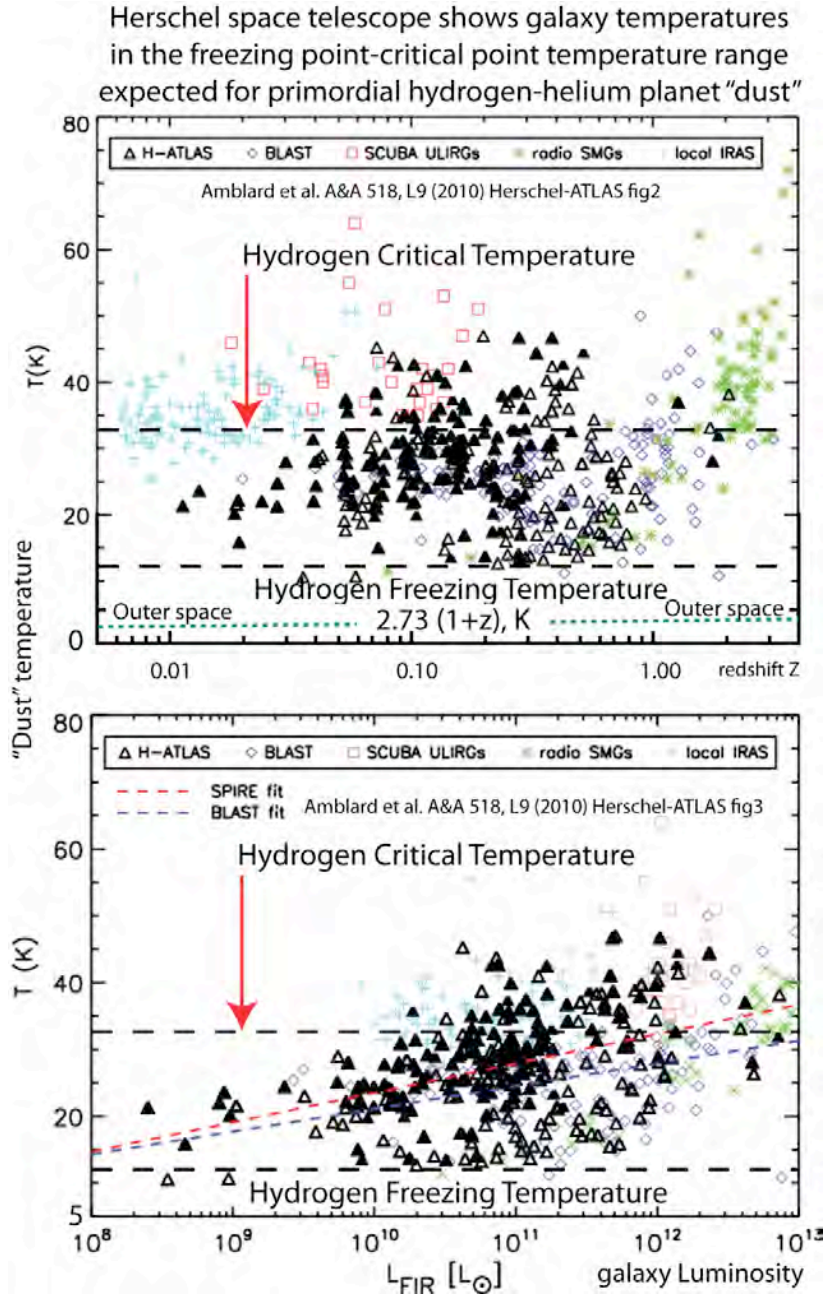

Fig. 5. Evidence from the Herschel-ATLAS temperatures that the "dust" of distant galaxies is frozen and heated primordial gas planets. The frozen hydrogen serves as a thermostat. Mergers of the planets to form stars increases the temperatures to above critical values, especially for highly luminous galaxies (bottom).

Fig. 5 (bottom) shows the temperature bounds of freezing-boiling hydrogen planets compared to the luminosity of the range of objects considered by Amblad et al. (2010) in their Herschel-ATLAS. Objects on the right with $10^{13}$ solar luminosity are elliptical galaxies. Objects on the left with only $10^8$ solar luminosity may be extremely dim galaxies whose brightness is determined by rare PGC clumps of planets that have formed stars.

Evidence that the dark matter of galaxies is primordial planets in clumps is provided by space telescopes sensitive to microwave and infrared frequency bands (IRAS, DIRBE, BOOMERang, WMAP). Maps of the sky show a "cirrus cloud" morphology of "dust", where the dust is at temperatures in the triple point to critical point range 14 K to 30 K to be expected if the cloud masses were dominated not by microscopic dust but by frozen hydrogen planets that thermostat the cloud temperatures (Gibson 2010, Gibson & Schild 2010). The same clouds appear at all wavelengths (13,000 to 100 $\mu$) and show signatures of bright PGC planet clumps with density $\sim$ 20 per squared degree. This gives a mass of $10^{42}$ kg for the Milky Way BDM halo assuming $10^{36}$ kg per PGC, confirming the HGD prediction that primordial planets in clumps dominate the mass of our Galaxy.

It is difficult to explain observations of multiple star and multiple massive planet systems using the standard models for star formation; that is, models not involving primordial planets as the source of all stars. As the resolution of the observations from space telescopes and ground telescopes increases over multiple frequency bands we see most stars are not alone but surrounded closely by large gas exo-planets and other stars. For example, the binary star system Upsilon Andromeda (http: // en.wikipedia.org / wiki / Upsilon_Andromedae) has two solar-mass stars and four confirmed (Curiel et al. 2011) 1-14 Jupiter-mass planets at distances from the stars of only 0.06 to 5 AU ($10^{10}$ m to $10^{12}$ m). It is easy to understand such a gravitational pile of Jupiters and stars formed at the center of a $10^{16}$ m diameter Oort cavity from HGD cosmology. Rather than multiple planets and multiple stars, one expects to get only one star per cloud and no planets if the cloud is gas and microscopic dust.

The primary star of the binary Upsilon Andromeda A (u And A) is close to its supernova II death from rapidly over eating planets, with mass 1.3 solar. Most systems of exo-planets include pulsars resulting from type II supernovae, which is the likely fate of u And A, McArthur et al. (2010). The secondary star of the binary (u And B) is $\sim 10^{14}$ m distant from u And A, still deep in the Oort cavity of the PGC.

## 4. Summary

We find no evidence to support the standard cosmological model based on condensation of cold dark matter CDM non-baryonic dark matter NBDM. The concept of CDM is unnecessary and physically impossible, as shown in Fig. 4. All gravitational condensations during the plasma epoch and during the gas epoch are dominated by the baryonic dark matter BDM and its viscosity. Viscous forces from photon viscosity $\nu \sim 10^{26}$ m$^2$ s$^{-1}$ permit gravity to fragment the plasma at mass scales decreasing from $10^{46}$ kg to $10^{42}$ kg in the time range $10^{12}$ s to $10^{13}$ s of the plasma epoch. Viscous speeds $\nu/L < V_S$ so sound speed $V_S$ and the Jeans scale $L_J$ are irrelevant for length scales $L < ct$ during this epoch. Pressure gradients along with pressure forces are rapidly erased by the passage of sound. The sonic peak in the CMB temperature anisotropy spectrum reflects the expansion at the plasma sound speed $c/3^{1/2}$ of proto-supercluster-voids as rarefaction waves. The sudden decrease at plasma to gas transition to $\nu \sim 10^{13}$ m$^2$ s$^{-1}$ of the 3000 K hydrogen-helium decreased the viscous fragmention mass from that of galaxies $10^{42}$ kg to that of Earth-mass planets $10^{25}$ kg.

The NBDM appears to be $\sim 97\%$ of the mass of the universe: some combination of neutrinos and possibly an additional neutrino-like weakly collisional material yet to be detected. The BDM is primordial gas planets in $\sim 10^{36}$ kg clumps with the density of globular star clusters $\sim 4\times10^{-17}$ kg m$^{-3}$, contrary to several microlensing consortia (MACHO, OGLE, EROS etc.) that claim to have excluded this form of BDM by finding an inadequate number of microlensing events, as shown in Figure 6 for the EROS programs (Renault et al. 1997). We suggest in Fig. 6 that these claims are questionable based on a flawed assumption that the planetary mass objects are not clumped and do not clump within clumps.

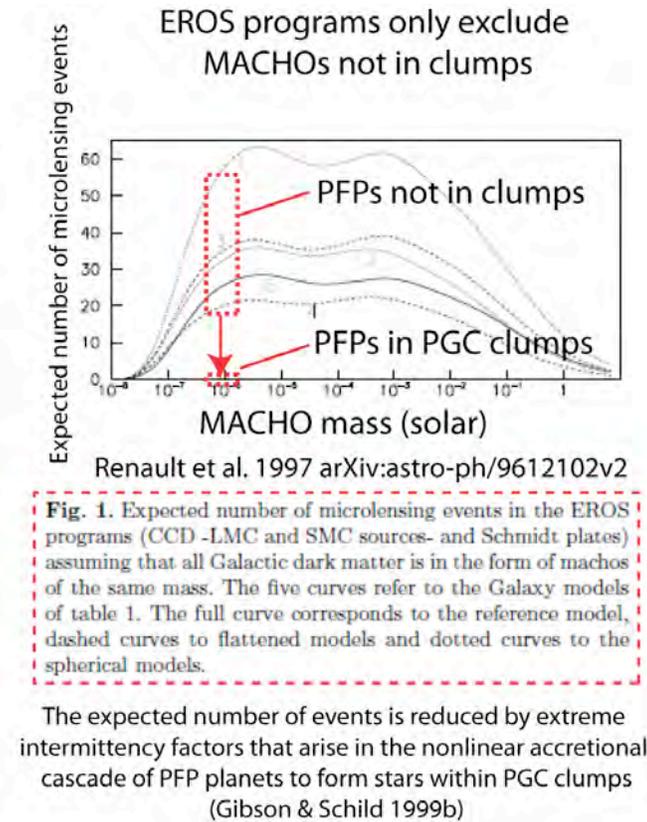

Fig. 6. EROS microlensing programs do not exclude primordial fog particle PFP planets with $10^{-6}$ solar mass as claimed (Renault et al. 1997 Fig. 1). A fatally flawed assumption is that such rogue planets are uniformly distributed rather than strongly clumped.

The collaborations all assume the planetary mass objects they exclude are uniformly distributed in the Galaxy halo, which is a strange assumption since such a massive population of planets must be frozen primordial gas planets since no other material besides hydrogen-helium is available in sufficient quantities, and such sticky gas planets would certainly form clumps whatever their initial conditions. As shown in Fig. 6, the expected number of microlensing events for PFP planets in starforming PGC clumps near the Galaxy PGC accretion disk is very small (Gibson & Schild 1999ab), but this is not considered in an EROS team review Lassere et al. (2000).

A flood of new evidence supports the HGD claim that the dark matter of galaxies is primordial planets in clumps that make all the stars. How else can hot Jupiters be explained, where Jupiter mass planets ($10^{27}$ kg) are found orbiting stars at Mercury orbits (0.03 AU). Recent examples are provided by Deming et al. (2011) and Desert et al. (2011). Figure 7 summarizes the hydrogravitational dynamics HGD model for gravitational structure formation.

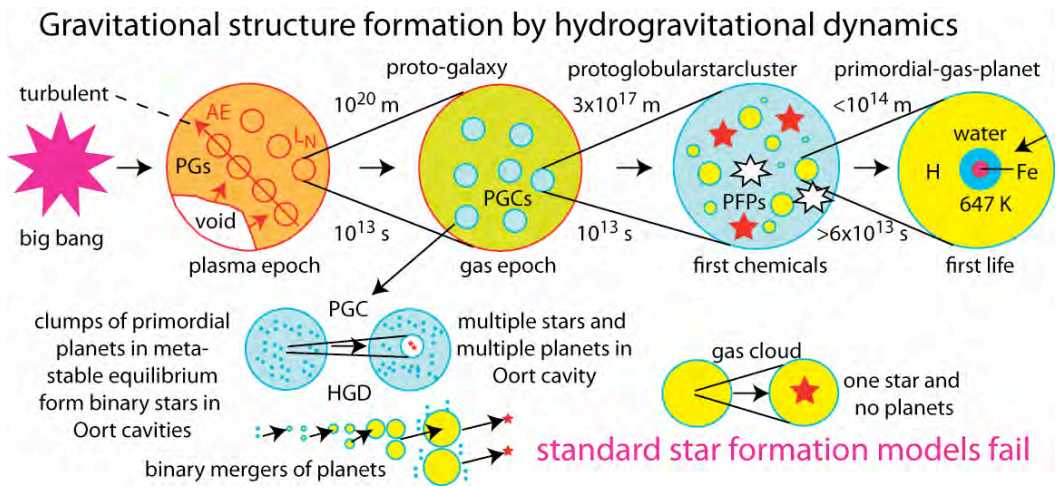

Fig. 7. HGD model for structure formation (top). HGD model for star formation compared to the standard model (bottom).

HGD length scales of Table 1 determine the various possible regimes of gravitational dynamics. Fig. 7 (top) shows plasma turbulence and superclustervoid fragmentation leading to protogalaxies in linear clusters. These fragment into Jeans scale clusters of primordial planets that form stars by binary mergers (bottom). The standard star formation models fail, and should be abandoned. Six gas planets on a flat orbit ~ $10^{10}$ m from a star have been recently observed by the Kepler Science Team (Lissauer et al. 2011). The system seems to be an accretion disk likely formed as gassy primordial planets merged to create the central star.

Figure 8 shows infrared and optical images of two carbon stars and the planetary nebula Helix, showing evaporated primordial planets in the process of star formation. Details about the carbon stars can be found in Olofsson et al. (2010), where the observed cometary globules are described as ejected from the star in spherical shells or formed by the interaction of ejected spherical shells. Such models are physically unrealistic and are rendered obsolete by observations and HGD cosmology. Huge density contrasts observed (and easily explained as accreting planet-comets) cannot form in the sun or be ejected, and could not form even if they were ejected. Further discussion of planet-comets in the Helix planetary nebula is given in Gibson, Wickramasinghe and Schild (2010, Fig. 4), where primordial planets are discussed as the sites for the first oceans and first life during a biological big bang that occurred only 2-8 Myr after the cosmological event.

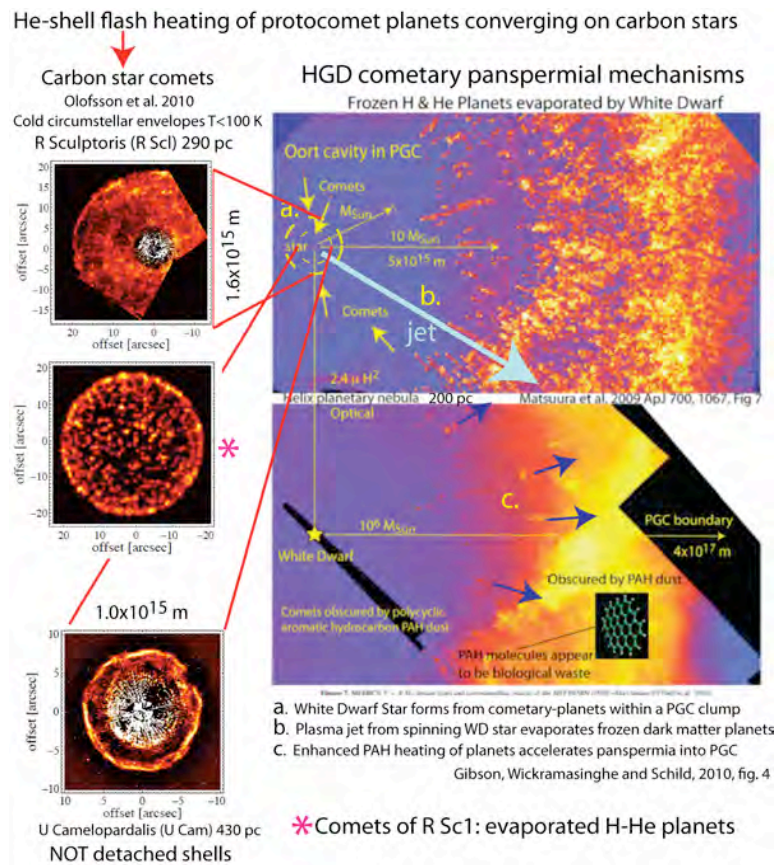

Fig. 8. Hubble space telescope images of carbon star comets (left) are interpreted as ambient primordial planets evaporated by a Helium shell flash. Sizes of the comet spheres are small compared to the Oort cavity of the Helix planetary nebula (right).

The carbon star images on the left of Fig. 8 are interpreted as examples of white dwarf and planetary nebula precursors, leading eventually to systems such as the Helix PNe, shown on the right of Fig. 8. The Helium flash of the carbon stars triggers a wave of frozen hydrogen planet evaporation that moves radially outward in a spherical shell that is not detached from the star but is formed in place. Some of the planet-comets will continue inward to feed the growing carbon star. Some of the gas evaporated will be expelled, slowing the star growth to form a white dwarf and eventually forming an Oort cavity in the PGC clump of planets.

The Herschel space telescope has detected an enormous water spray emerging from carbon star He-flash events, with rates ($4 \times 10^{26}$ kg/yr) up to a million earth-oceans per year in a sample of eight C-stars (Neufeld et al. 2010). Such a spray suggests and supports evaporation of large populations of ambient primordial planet-comets and their frozen water oceans by AGB He-shell thermal pulses from the carbon stars.

Figure 9 shows a collection of transiting small exoplanets detected by the Kepler Science Team (Lissauer et al. 2011). Planets very close to their star (b, c, d, e, f) have temperature ~ 647 K, the critical temperature of water, suggesting their water oceans are being evaporated as the primordial planet water oceans exceed the water critical temperature 647 K.

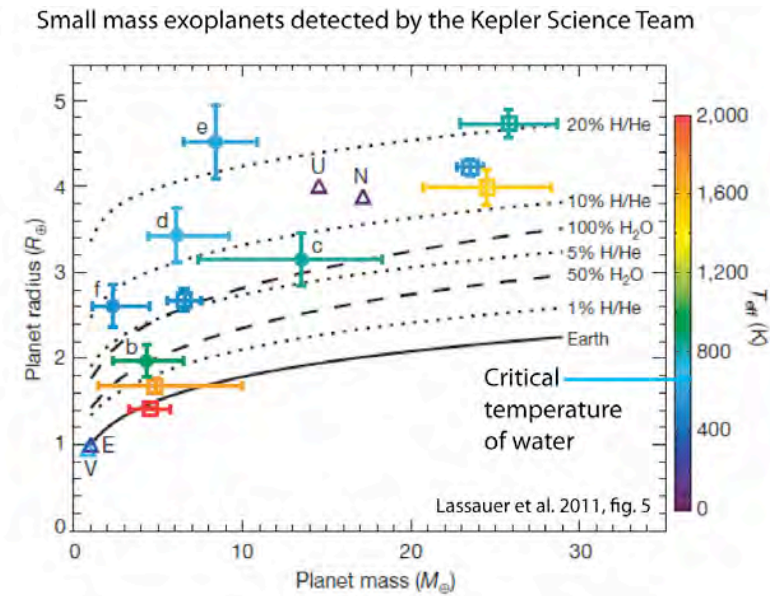

Fig. 9. Small mass exoplanets b, c, d, e detected by the Kepler Science Team are ~ $10^{10}$ m from star Kepler-11 on tight orbital planes within 0.5° of 89°. An accretion disk of primordial planets with evaporating water oceans is indicated. Solar planets E, V, U, and N are shown as triangles. Other small Kepler exoplanets are shown by open squares (Lassauer et al. 2011, fig. 5).

A crisis concerning mass transfer in star formation is termed the 3He (tralphium, tralpha particle) problem[1]. This isotope is produced in trace quantities by big bang nucleosynthesis (Fuller and Smith 2011), and in brown dwarfs by burning tritium. However it is also produced in stars, but is not observed in the large quantities expected if stars eject their envelopes, as shown in Figure 10.

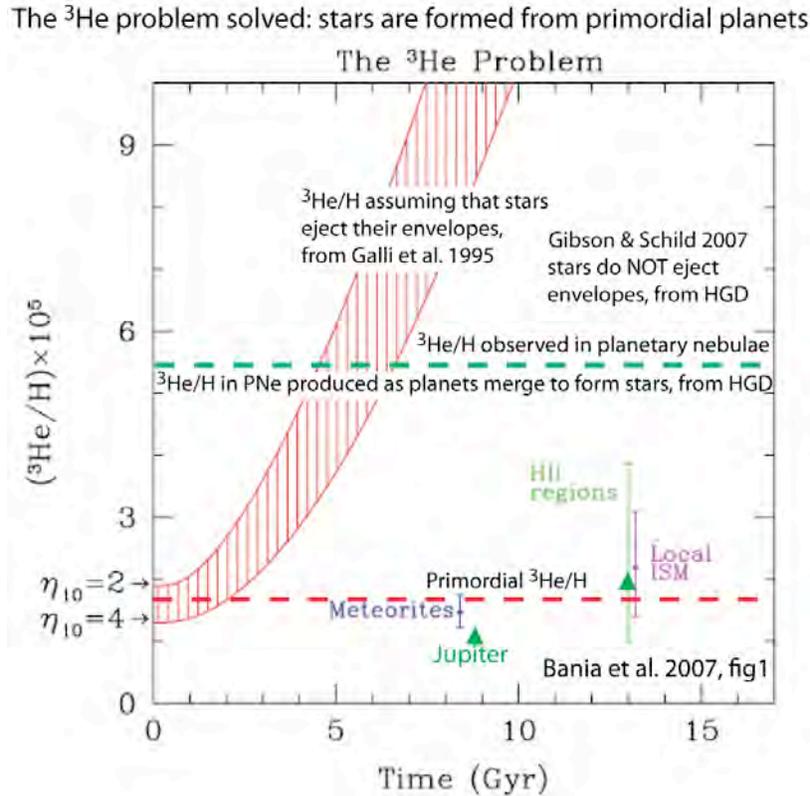

Fig. 10. Solution of the $^3$He problem (Bania et al. 2007) is provided by hydrogravitational dynamics HGD, where all stars are formed by primordial planet mergers, and do NOT eject stellar materials in massive shells as claimed by the standard models of star formation.

Primordial values of $^3$He/H ~ $10^{-5}$ are observed in ionized, Oort cavity size HII regions, Leto et al. (2009). This is expected from HGD cosmology (Nieuwenhuizen 2011) because comet cores are evaporating primordial planets, with primordial gas abundances, and because it is a myth that stars eject massive envelopes by superwinds during their red giant phase, Gibson and Schild (2007). The increase in

---

[1] *And God said: "Let there be mass three." And there was mass three. And God saw tritium and tralphium, and they were good".* George Gamov about 1949.

$^3$He/H shown by the red shaded region of Fig. 10 is NOT reflected near stars, as shown by the primordial $^3$He/H values in HII regions, Bannia et al. (2007). The large $^3$He/H values detected in planetary nebulae PNe can be explained by plasma jet pumping by the central white dwarf. Similar large $^3$He/H ratios are detected in old globular clusters, presumably for the same reason.

## 5. Conclusions

The dark matter of galaxies is mostly frozen primordial planets PFPs in proto-globular-star-cluster PGC clumps, as predicted by hydrogravitational dynamics HGD cosmology. All stars are formed within PGCs by mergers of PFPs. Increasingly precise observations by microlensing and by infrared and microwave telescopes support the predictions of HGD cosmology. Microlensing collaborations such as EROS and MACHO have not excluded planetary mass objects in clumps as the dark matter of galaxies. The observations should be repeated with much higher sampling frequencies in directions that include foreground PGC clumps of primordial planets.

The standard ΛCDMHC cosmology is physically unrealistic (Fig. 4) and increasingly in conflict with observation (Fig. 2) and basic fluid mechanical constraints. It should be abandoned. HGD cosmology corrects the standard model for many unrealistic fluid mechanical assumptions (Figs. 3) and is abundantly supported by observations (Figs. 1,2,5,8,9,10). Microlensing observations that exclude primordial planets in clumps as the galaxy dark matter (Fig. 6) must be re-interpreted. Standard models for star formation based on ΛCDMHC cosmology must be abandoned (Fig. 7). A flood of new observational evidence of exoplanets further supports the HGD claim that all stars form by mergers of primordial planets, and the more recent HGD claim that life would probably not have happened nor have been widely transmitted within and between galaxies without the millions of primordial planets per star, as predicted by HGD cosmology. HGD should be adopted as the new standard cosmological model.

As shown in Fig. 10, the primordial $^3$He/H concentration $\sim 10^{-5}$ observed in HII regions strongly supports the HGD star formation paradigm (Niewenhuizen 2011).